\documentclass[proof]{pasj00}
\draft

\begin{document}
\SetRunningHead{Y.~S.~Honda and M.~Honda}
{Energetic Neutrino Production in Mrk 501}

\Received{2009/07/19}
\Accepted{2010/4/6}

\title{Additional Acceleration of Protons 
and Energetic Neutrino Production \\
in a Filamentary Jet of the Blazar Markarian 501}

\author{Yasuko S. \textsc{Honda}}%
\affil{Department of Total Systems Engineering, 
Kinki University Technical College, \\ Kumano, Mie 519-4395}
\email{yasuko@ktc.ac.jp}

\author{Mitsuru \textsc{Honda}}
\affil{Plasma Astrophysics Laboratory, Institute for Global Science, 
Kumano, Mie 519-4327}


\KeyWords{galaxies: individual (Markarian 501) --- magnetic fields --- 
methods: numerical --- neutrinos: individual (Markarian 501) --- shock waves} 

\maketitle

\begin{abstract}
Blazars have been regarded as one of the most powerful sources of 
the highest energy cosmic rays and also their byproducts, neutrinos. 
Provided that a magnetized filamentary system is established in a 
blazar jet as well, we could apply the mechanism of multi-stage 
diffusive shock acceleration to a feasible TeV emitter, Mrk 501  
to evaluate the achievable maximum energy of protons. Taking 
conceivable energy restriction into account systematically, it seems 
adequate to say that EeV-protons are produced at this site by our 
present model. We also estimate neutrino fluxes generated by these 
accelerated protons and discuss the detectability based on an updated 
kilometre-scale telescope such as IceCube. 
\end{abstract}

\section{Introduction}
The origin of the ultra-high energy cosmic rays (UHECRs) beyond 
100 EeV has still been an enigma in modern astrophysics. UHECRs are 
frequently accompanied by significant fluxes of energetic neutrinos 
generated in processes associated with hadronic interactions. 
Although the source direction of cosmic rays with energy above 60 
EeV can be identified by an analysis of recent Auger data 
(Abraham et al. 2007), neutrinos are regarded as being the most useful probe 
of their source confirmation owing to no deflection by 
galactic/intergalactic magnetic fields and the weakest reduction by 
interactions with the cosmic background radiation. The models for 
the origin of the highest energy cosmic rays are classified 
into two categories: top-down and bottom-up. In top-down 
scenarios, neutrinos are presumed to be decay or annihilation 
products of cosmological remnants with the Grand Unified energy 
scale of $\sim 10^{15}$ GeV. Within this framework, numerous 
models have been proposed so far, such as topological defects, 
Z-bursts, annihilation of dark matter particles and so on 
(for a review, see \cite{bhattacharjee00}). In an alternative 
approach, the bottom-up model is based on the idea that the 
highest energy cosmic rays are generated by astrophysical 
objects. The most plausible candidates are gamma ray bursts 
(GRB: \cite{waxman97,vietri98}) and active galactic nuclei 
(AGN: \cite{honda04a,honda09}). In addition to these objects, 
it is pointed out that UHECRs are also produced via  
stochastic acceleration in the giant lobes of radio galaxy 
(Cen A: \cite{fraschetti08,sullivan09}), which accounts for 
the past year discovery by the HiRes and Auger collaborations. 
Supernova remnants, X-ray binaries, 
mini-quasars (e.g.\ \cite{gaisser95}, for a review), and 
soft gamma ray repeaters (SGR: \cite{halzen05a,ioka05})
can also be stellar type neutrino sources, any of which could 
produce observable fluxes of energetic neutrinos. If protons are 
accelerated at these astrophysical objects, neutrinos are expected 
to be produced in collisions with ambient photons or protons. 

Some of the above-mentioned models predict neutrino flux at the 
level of a few events per km$^{2}$ year. The upper range of this 
estimation seems to be within reach of a first-generation 
neutrino telescope, such as AMANDA (Antarctic Muon And Neutrinos 
Detector Array). Among a wide variety of physics topics to be 
explored with the neutrino telescope, the most important goal is to 
search for the origin of cosmic radiation, especially that originated 
from AGN and/or GRB. For this purpose, AMANDA has been primarily 
optimised in the energy range from TeV to PeV, targeting not 
only the diffuse, but point source, flux of energetic neutrinos. 
According to an analysis of AMANDA data, we unfortunately have 
no evidence for point sources so far, (Woschnagg et al. 2005; 
Abbasi et al. 2009), 
except for a temporal coincidence with an orphan flare of the TeV 
blazar, 1ES 1959+650 (Halzen \& Hooper 2005). The IceCube, a cubic 
kilometre-scale neutrino telescope, which is partly operating and 
still under construction as a successor to AMANDA at the same site, 
is designed to detect the fluxes about 10-50 events per km$^{2}$ year.
Similar event rates are predicted by model calculations assuming 
that AGN or GRB are the actual source (Halzen 2005) and hence  
the IceCube is expected to certainly detect neutrino fluxes from such  
point sources. In order to observe a {\it guaranteed} source, 
however, even the km$^{3}$-sized detector should safely operate over 
a period of at least ten years. 
   
As for the cosmic accelerators, the most promising mechanism to lead highest 
energies and power-law energy spectrum is considered to be of 
diffusive shock acceleration (DSA: \cite{drury83,lagage83a,lagage83b}). 
Solar energetic particles and galactic cosmic rays are consistently 
explained by this mechanism applied to shocks at heliosphere and 
supernova remnants, respectively (for a review, see \cite{blandford87});   
these acceleration sites have actually been confirmed by some 
observations (Koyama et al. 1995). 
Polarization data in the range of radio to optical wavelengths imply 
that {\it in situ} acceleration of electrons (and possibly ions) must 
also be taking place in the knots of AGN jets (Honda \& Honda 2004a) or  
in the hot spots of Fanaroff-Riley type-II (FR-II) sources 
(probably terminal shocks) (Perley et al. 1984; Carilli \& Barthel 1996). 
At these sites, magnetic fields in the vicinity of a shock front 
plays an essential role in particle acceleration. A detailed configuration 
of the fields in extragalactic jets has been revealed by polarization 
measurements using very long-baseline interferometry.    
For example, the quite smooth fields predominantly transverse to the 
jet axis, are typically observed in the core region of BL Lac objects 
(1803+784: \cite{gabuzda99}; 0300+470: \cite{nan99}). Allowing the 
propagation of a shock wave through the jets, this implies the 
establishment of a (quasi-) perpendicular shock, which is capable of  
accelerating cosmic ray particles efficiently. Evidence for 
large-scale toroidal magnetic fields has been also 
discovered in the galactic center (GC) region (Novak et al. 2003). 
Based on the above results, it seems adequate that system 
of magnetized filaments has been established in blazar jets as well. 

We have suggested a theoretical model to account for the generation of 
such a large-scale toroidal (transverse) magnetic field in 
astrophysical jets (Honda \& Honda 2002). 
In this model, huge currents launched from a central engine are broken 
into many filaments whose transverse sizes are self-adjusted for the 
electromagnetic current filamentation instability (CFI: \cite{honda04}
and references therein). In the nonlinear stage of the CFI, the 
magnetized filaments are often regarded as being strong turbulence 
that can largely deflect charged particles. Allowing shock 
propagation through the jet, the particles are expected to be quite 
efficiently accelerated for the DSA scenario, which appears to be 
favorably taking place in the AGN jets. Indeed, some knots of a 
radio galaxy jet are associated with the shock fronts (M87: 
\cite{biretta83,capetti97}) and circumstantial evidence for in-situ 
acceleration of electrons (Meisenheimer et al. 1989) have been found. 
It was also pointed that a similar pattern of small-scale quasi-static 
magnetic fields can also be established by some numerical simulations: 
e.g., during the collision of electron-positron plasmas existing 
in SNRs, pulsar winds, GRBs, relativistic jets, and so on 
(Kazimura et al. 1998; Silva et al. 2003). 
Using a three-dimensional relativistic electromagnetic code, 
Nishikawa (2003) show that non-uniform small-scale magnetic field is 
generated due to the Weibel instability at a jet front propagating 
through an ambient plasma with/without initial fields. 
These dynamics might be involved in some of the knots in the FR-I 
radio jets, which appear to be a shock established when faster material 
is overtaking with slower one (M87: \cite{biretta83}).

In the case of blazars, however, the detailed internal structure of 
their jets has still not been confirmed because of optical thickness 
with respect to the observer's line of sight. Another approach to shed 
light on their configuration is provided by the remarkable short variability 
timescale of a blazar light curve (e.g., Mrk 421: reaching a few 
minutes, \cite{cui04,blazejowski05}). This is significantly shorter than the 
light-crossing time at the black hole horizon, which implies the presence of  
some substructure in the parsec-scale jet. According to our model 
configuration of the filamentary jet, the strength of the local magnetic 
field is described by a power-law of the filaments transverse sizes.  
If charged particles are injected into this system, they are 
diffusively accelerated by a collisionless shock being scattered by 
the field fluctuations. Since the efficiency of the acceleration and 
loss depends upon the spatial size scales, the local maximum 
energies of accelerated particles are also characterized by their 
nearest filament sizes. 
The spectrum extending to the X-ray region is attributed to the 
synchrotron radiation of accelerated electrons. 
In particular, the correlation between X-ray and gamma-ray light curves 
of Mrk 421 is well reproduced by a model of the hierachical turbulent 
structure of the jet (Honda 2008).
The most interesting consequence is that the transition 
of a hierachical turbulent structure seems to be responsible for the observed 
patterns of energy-dependent light curves, e.g., soft lag (Takahashi 
et al. 1996; Rebillot et al. 2006) or hard lag (Fossati et al. 2000) 
and a tight-correlation mode (Sembay et al. 2002).  

As for Mrk 501, which is also firmly established as a TeV 
$\gamma$-ray emitter, the X-ray light curve shows a very rapid flare 
varying over several 100 seconds (Catanese \& Sambruna 2000). It is 
also revealed that the TeV flares correlate with X-ray radiation on 
timescales of hours or less from multiwavelengths observations 
(Pian et al. 1998; Catanese et al. 1997; Krawczynski et al. 2000; 
Sambruna et al. 2000). Moreover, the largest shift of the peak energy 
during the peak-luminosity change was observed among all blazars 
(Kataoka et al. 2001). Summarizing these observational results, we can say 
that the acceleration of electrons similar to Mrk 421 is also taking 
place in Mrk 501. Presuming a finite ion abundance in the jet of Mrk 501 
(Rawlings \& Saunders 1991), 
such a DSA mechanism operates for arbitrary nuclei (of course including 
protons) as well. Indeed, from the X-ray spectrum 
of SS 433 jet, emission lines of various elements such as Ne, Mg, Si, and 
Fe were observed (Kotani et al. 1996). 

In the present paper, we evaluate the achievable maximum energy of protons
in the Mrk 501 jet to estimate the resultant neutrino fluxes in accordance 
with our diffusion and acceleration mechanism in a filamentary jet, which 
we have proposed and developed in a series of papers 
(Honda \& Honda 2004b; 2005; 2007). 
In section 2, installing our model structure of magnetized 
filaments to the jet of Mrk 501, we describe the acceleration 
proccesses of protons using our novel DSA mechanism.  
Taking competitive energy losses and restrictions into consideration, we 
present the scaling of the maximum energy. 
In \S 2.1 we describe the acceleration taking place at local magnetic 
fields induced by the current filament each (referred to as preliminary 
acceleration) and in \S 2.2 the additional acceleration due to the 
inter-filaments' deflection. In \S 2.3, we calculate the 
achievable maximum energies of accelerated protons with respect to the 
transverse filament sizes and magnetic field parameters. 
We then evaluate neutrino flux from Mrk 501 and discuss its detectability 
based on a km-scale neutrino telescope such as IceCube  
in section 3. Conclusions are summarized in section 4. 

\section{Proton Acceleration in Magnetized Filaments}
Suppose a parsec-scale blazar jet transporting energetic particles 
from the central engine. The directional plasma flow will favorably 
induce huge currents (Appl \& Camenzind 1992) that breaks up into 
many filaments because of electromagnetic CFI.
Each filament generates a small-scale transverse magnetic field whose 
strength is determined by the transverse size of a filament,  
$\lambda$, i.e., $|\mbox{\boldmath$B$}|=B_{\rm m}(\lambda/D)
^{(\beta-1)/2}$, where $B_{\rm m}=|\mbox{\boldmath$B$}|_{\lambda=D}$; 
$D$ and $\beta=4.3$ denote the jet 
diameter and power-law spectral index of the magnetized filamentary 
turbulence, respectively (Honda \& Honda 2007). 
Viewing the jet globally from outside, randomly oriented fields inside 
jet are cancelled, except for a large-scale toroidal magnetic field along 
the envelope. 

As was illustrated in our previous paper (Honda \& Honda 2005), there exists 
an energy hierarchy for protons trapped in such a filamentary current 
system: (i) $E_{\rm p}\ll |eA|$ and (ii) $E_{\rm p}\gg |eA|$, where $e$ 
and $A$ are the charge of an electron and the vector potential, respectively. 
The former corresponds to a low-energy regime, in which protons are 
strictly bounded for the local magnetic field induced by each filament. 
The latter reflects the higher energy regime in which the validity 
condition of the quasilinear approximation, 
$\langle f_{\rm p}\rangle \gg|\delta f_{\rm p}|$ is satisfied, 
where $\langle f_{\rm p}\rangle$ 
and $\delta f_{\rm p}$ are the averaged and fluctuated part of the 
momentum distribution function for protons in the test-particle 
approximation, $f_{\rm p}$, respectively. In this regime, protons 
are no longer trapped with local magnetic fields, but 
are deviated with small fluctuations. Below, we describe the 
acceleration and competitive energy losses in each energy hierarchy.      

\subsection{Preliminary Acceleration} 
In the low-energy regime of $E_{\rm p}\ll|eA|$, protons 
are strictly bounded 
and gyrating around the local magnetic field induced by each filament. 
Energies of injected protons are elevated via the conventional 
diffusive shock acceleration being resonantly scattered from small 
magnetic fluctuations (Drury 1983). The characteristic acceleration 
time of protons is described by 
\begin{equation}
t_{\rm pre,acc}\simeq \left(\frac{3\eta_{\rm p}r_{\rm g,p}}{c}\right)
\left(\frac{r-1}{r}\right),
\end{equation}
where $\eta_{\rm p}=(3/2b)(\lambda/2r_{\rm g,p})^{2/3}$, presuming  
Kolmogorov turbulence, 
$b$ ($\ll 1$) is the energy density ratio of fluctuating 
to local mean magnetic fields, $r_{\rm g,p}$ is the proton gyroradius, 
and $r$ (=4 for the strong shock limit) is the shock compression ratio.  
The achievable maximum energies of protons are limited by the various 
radiative cooling timescales, and here we consider two representatives: 
the synchrotron radiation losses, $t_{\rm p, syn}$, and collisions with 
ambient photons, $t_{\rm p\gamma}$. In blazar jets the propagation time 
of shock, $t_{\rm sh}$, also restricts the highest energy, and hence  
the temporal limit is given by 
\begin{equation}
t_{\rm pre,acc}=\min(t_{\rm p\gamma}, t_{\rm p,syn}, t_{\rm sh}).
\end{equation}
Besides the temporal limit, there exists a spatial limit: the maximum 
gyroradii of protons are not allowed to exceed the transverse sizes 
of in-situ filaments. The achievable energy of protons via the preliminary 
acceleration is determined by comparing the temporal and spatial limit. 
In the following, conceivable energy constraints for Mrk 501 are provided. 

\subsubsection{Synchrotron cooling losses}
Gyrating around the filament-induced local fields, and being scattered by 
small fluctuations, protons emit synchrotron photons, which can be a dominant 
cooling process when the energy density of the magnetic field is sufficiently 
greater than that of the radiation. The characteristic loss 
time for proton synchrotron is given by (Rybicki \& Lightman 1979) 
\begin{equation}
t_{\rm p,syn} = \frac{3m_{\rm p}c}{4\sigma_{\rm T}u_{\rm B}
\gamma_{\rm p}} \left(\frac{m_{\rm p}}{m_{\rm e}}\right)^{2}
\simeq 1.2\times 10^{13}\left(\frac{0.02~{\rm G}}{B}\right)^{2}
\left(\frac{10^{9}}{\gamma_{\rm p}}\right)~{\rm s},
\end{equation}
where $m_{\rm p}$ and $m_{\rm e}$ are the masses of a proton and of 
an electron, respectively; $\sigma_{\rm T}$ is the cross section for 
Thomson scattering and $\gamma_{\rm p}$ is the Lorentz factor of 
accelerated protons. The average energy density of the local magnetic 
field is denoted by $u_{\rm B}=B^{2}/(8\pi)$, where $B=|\mbox{\boldmath$B$}|$.
It should be noted that the adopted field value of $B_{\rm m}=0.02$ G 
(Tavecchio \& Maraschi 2001) is not the one averaged over the compact blob but 
the maximum for a filament whose radial size is compared to 
the jet diameter.    

\subsubsection{Collisions with ambient photons}
We consider proton-photon collisions leading to a  
pion-production cascade. Since the collision timescale is characterized by 
the target photon spectrum, which is still unknown, we adopt here a 
description by a single power-law: $n(\varepsilon_{\gamma})
\propto \varepsilon_{\gamma}^{-2}$ (\cite{bezler84}), where 
$n(\varepsilon_{\gamma})$ is the number density of photons per unit 
energy interval $d\varepsilon_{\gamma}$. Then, the timescale can be 
expressed as 
\begin{equation}
t_{\rm p\gamma}=\chi^{-1}\left(\frac{u_{\rm B}}{u_{\gamma}}\right)
t_{\rm p,syn}\simeq 6.0\times 10^{10}
\left(\frac{10^{9}}{\gamma_{\rm p}}\right)~{\rm s},
\end{equation}
where $u_{\gamma}\simeq 4.0\times 10^{-4}$ ergcm$^{-3}$ 
(\cite{bicknell01}) is the average energy density of target photon 
fields and $\chi\simeq 200$ (\cite{biermann87}). Since $u_{\rm B}/u_{\gamma}
\simeq 10^{-2}$, it is found that $t_{\rm p\gamma}\ll t_{\rm p,syn}$. 

We should also check whether collisions with particles (especially, 
with protons) becomes effective or not. The characteristic cooling time 
of relativistic protons due to inelastic pp collisions can be written 
as (\cite{aharonian04})
\begin{equation}
t_{\rm pp}=\frac{1}{n_{0}\sigma_{\rm pp}fc}\simeq 1.7\times 10^{15}
\left(\frac{1~{\rm cm}^{-3}}{n_{0}}\right)
\left(\frac{40~{\rm mb}}{\sigma_{\rm pp}}\right)~{\rm s}, 
\end{equation}
where the number density of the target hydrogen medium, $n_{0}\simeq 1$ 
cm$^{-3}$, corresponds to the upper limit assuming mass loaded 
hadronic jet models. The averaged total cross section, $\sigma_{\rm pp}$,  
is approximately 40 mb at very high energies, and $f\simeq 0.5$ is the 
coefficient of inelasticity. It would be fair to say that collisions 
with protons should not be taken into consideration. 

\subsubsection{Propagation time of shock}
The achievable maximum energy of protons is also restricted by the duration 
that the shock is propagating through the jet from the central engine to the 
working surface. This corresponds to the age of a blob, which is 
propagating with mild-relativistic speed through the relativistic jet 
(\cite{honda04a}), and is approximately written by $L/U$.  
For Mrk 501, we obtain  
\begin{equation}
t_{\rm sh}=\frac{L}{U}\simeq 1.5\times 10^{9}
\left(\frac{L}{7.4~{\rm pc}}\right)\left(\frac{0.5c}{U}\right)~ {\rm s},  
\end{equation}
where $L\simeq 7.4$ pc is the distance from the core to the blob 
(\cite{giovannini99}) and $U\simeq 0.5c$  is the average speed of the 
blob (\cite{muecke01}). 
This blob is currently operating for particle acceleration, and 
therefore $t_{\rm sh}$ cannot be compared to the adiabatic loss  
timescale for AGN, which corresponds to the lifetime of the shock 
(\cite{muecke03}). The timescale of adiabatic losses is expressed 
as $t_{\rm ad}=3L/(2\Gamma
_{\rm J}v_{\rm r})$, where $\Gamma_{\rm J}$ and $v_{\rm r}$ are the 
Lorentz factor of the jet and the speed of the radial expansion. Considering 
narrow opening angles of blazars, which implies $v_{\rm r}\ll U$, we can 
safely say that $t_{\rm sh}$ is sufficiently shorter than $t_{\rm ad}$.   
 
\subsubsection{Escape from local filaments}
Once accelerated particles acquire sufficient energies, they are 
escaping from the local magnetic fields, given as a function of the  
filament sizes. This spatial limit is given by the assumption that 
the gyroradius of each particle should not be beyond the transverse 
size of the local filament. Then, the local maximum energy is denoted by 
$E_{\rm p}^{*}=eBr_{\rm g,p}\leq eB\lambda/2$.  

The transverse sizes of the minimum and maximum filaments are regarded 
as being comparable to the Debye sheath and the jet diameter, respectively. 
Thus, the energy of particles escaping from the largest filament can be 
written as 
\begin{equation}
E_{\rm p}^{*}=\frac{eB\lambda}{2}\leq\frac{eB_{\rm m}D}{2}\simeq 6.2
\times10^{17} \left(\frac{B}{0.02~{\rm G}}\right)
\left(\frac{D}{0.067~{\rm pc}}\right)~{\rm eV}, 
\end{equation}
where $D\simeq 0.067$ pc is the extent of the radio emitting region 
(\cite{giroletti08}). This value is estimated by substituting 
$0.02$ G for $B$ in the formula suggested by Marscher (1987). 

\subsection{Additional Acceleration} 
Once protons are accelerated and injected into the high-energy regime 
of $E_{\rm p}\gg|eA|$, they are no longer bounded to the local magnetic 
field. Field vectors are randomly oriented in the internal jet in the 
transverse direction to the filament, and hence these protons can move 
almost freely in the jet being deflected by the random magnetic field. 
Allowing for shock propagation, protons are additionally accelerated in 
this regime, being off-resonantly scattered in a forest of magnetized 
filaments. In this aspect, the preliminary accelerator via the 
conventional DSA mechanism can be regarded as being an 'injector' to this 
further booster. The new injection mechanism is referred to as 
transition injection, in analogy to bound-free transition in atomic 
excitation. The characteristic time of the additional acceleration for 
protons is given by 
\begin{equation}
t_{\rm add,acc}=\frac{3\sqrt{6}\pi\beta r(r+1)}{8(\beta-1)(\beta_{\rm p}+1)
(\beta_{\rm p}+2)(r-1)}\frac{cE_{\rm p}^{2}}{e^{2}B_{\rm eff}^{2}DU^{2}}, 
\end{equation}
where $\beta_{\rm p}=3$ (e.g., \cite{stecker99,demarco03}) and 
$E_{\rm p}=\gamma_{\rm p}m_{\rm p}c^{2}$ are the spectral index and 
energy of protons, respectively; $\beta$ is the filamentary turbulent 
spectral index, and $|B_{\rm eff}|^{2}/B_{\rm m}^{2}\sim 0.5$ is 
assumed. The maximum energy boosted by the additional acceleration 
is also restricted by some radiative loss processes. In addition to 
two possible timescales: for diffuse synchrotron, $t_{\rm d,syn}$, and 
for the p$\gamma$ collision, $t_{\rm d,p\gamma}$, we consider the shock 
propagation timescale, $t_{\rm sh}$, and therefore  
$t_{\rm add,acc}=\min(t_{\rm d,syn}, t_{\rm d,p\gamma}, t_{\rm sh})$.  
The spatial limit will also be considered. 

\subsubsection{Diffuse synchrotron losses}
The additionally accelerated protons deflected by the 
random magnetic fields emit unpolarized synchrotron radiation. 
Here, we adopt the characteristic cooling time for the diffuse 
synchrotron from our previous paper (\cite{honda07}), which was 
derived from the theoretical basis by \citet{toptygin87}: 
\begin{equation}
t_{\rm d,syn}=\frac{3m_{\rm p}c}{4\sigma_{\rm T}u_{\rm B}\gamma_{\rm p}}
\left(\frac{m_{\rm p}}{m_{\rm e}}\right)^{2}\frac{1}{36\pi^{2}}
\frac{\beta(\beta+2)^{2}(\beta+3)}{2^{\beta}(\beta^{2}+7\beta+8)}
=t_{\rm p,syn}\tilde{\tau}(\beta),  
\end{equation}  
where $t_{\rm p,syn}$ is the timescale of normal synchrotron losses 
and $\tilde{\tau}(\beta)\equiv(6\pi)^{-2}[\beta(\beta+2)^{2}
(\beta+3)/2^{\beta}(\beta^{2}+7\beta+8)]$. For $\beta=4.3$ (\cite{honda07}),  
the maximum energy of a proton,  
$\gamma_{\rm p}$, is directly estimated by the equation 
$t_{\rm add,acc}=t_{\rm d,syn}$, which gives 
\begin{equation} 
\gamma_{\rm p}\simeq 2.7\times 10^{11}\left(\frac{U}{0.5c}\right)^{2/3}
\left(\frac{D}{0.067~{\rm pc}}\right)^{1/3}.
\end{equation} 
It should be noted that the maximum energy restricted by the diffuse 
synchrotron is free from the magnetic field strength because of the 
same $B$-dependence of $t_{\rm add,acc}$ and $t_{\rm d,syn}$. 

\subsubsection{Energy losses due to p$\gamma$ collision}
Cooling due to collisions with photons is also characterized by the 
above-mentioned diffuse synchrotron: 
$t_{\rm d,p\gamma}=\chi^{-1}(u_{\rm B}/u_{\gamma})t_{\rm d,syn}$. 
Using the same target photon spectrum and other notation with those 
given in $\S 2.1.2$, we obtain 
\begin{equation} 
\gamma_{\rm p}\simeq 4.6\times 10^{10}\left(\frac{U}{0.5c}\right)^{2/3}
\left(\frac{D}{0.067~{\rm pc}}\right)^{1/3}.
\end{equation} 

\subsubsection{Restriction from the Age of the Blob}
As we have mentioned in \S 2.1.3, the shock propagation time regarded as 
the age of the blob restricts the acceleration. Equating 
$t_{\rm add,acc}=t_{\rm sh}$, we obtain  
\begin{equation} 
\gamma_{\rm p}\simeq 4.3\times 10^{9}\left(\frac{U}{0.5c}\right)^{1/2}
\left(\frac{L}{7.4~{\rm pc}}\right)^{1/2}\left(\frac{D}
{0.067~{\rm pc}}\right)^{1/2}. 
\end{equation} 

\subsubsection{Spatial Limit for Additional Acceleration}
It should be noted that additionally accelerated particles are not 
gyrating to the local fields, but deflected by randomly oriented fields. 
Thus, the spatial limit is determined by the condition that the mean free 
path for this collisionless deflection should be smaller than the 
system size: $\ell\sim\tilde{\rho}D$, where $\tilde{\rho}>1$ is a 
dimensionless parameter (see, \cite{honda05}, for the details). It gives 
\begin{eqnarray}
E_{\rm p}^{\rm max}&=&\sqrt{\frac{2(\beta-1)(\beta_{\rm p}+2)
(\beta_{\rm p}+4)\psi_{1}^{2}}
{\pi\beta}}\tilde{\rho}
^{1/2}eB_{\rm eff}D \nonumber \\
&\simeq&2.1\times 10^{18}\left(\frac{B_{\rm eff}}
{0.014~{\rm G}}\right)
\left(\frac{D}{0.067~{\rm pc}}\right)~{\rm eV},
\end{eqnarray}
where $\tilde{\rho}=1$ is assumed.

\subsection{Achievable Maximum Energy of Protons}
\begin{figure}
  \begin{center}
    \FigureFile(80mm,80mm){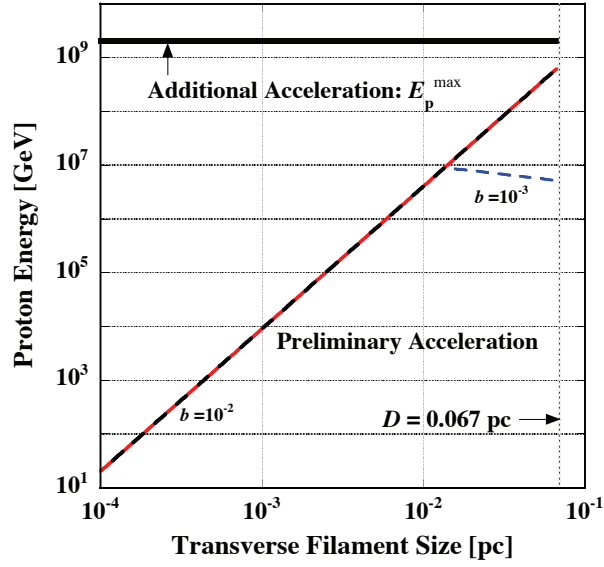}
  \end{center}
  \caption{Maximum energies of accelerated protons versus the transverse 
filament sizes for various $b$-values. The value of $B_{\rm m}$ is fixed 
at 0.02 G. The monotonically increasing line and its branch denote the 
maximum energies achieved by the preliminary acceleration. Solid and 
dashed lines correspond to the cases of $b=10^{-2}$ and $10^{-3}$, 
respectively. The horizontal bold line denotes the maximum energy 
achieved by additional acceleration.}
\label{fig:fig1}
\end{figure}

Putting together the above discussion, provided both the normal and 
transition injection of protons are realized in a blazar jet, 
multi-stage diffusive shock acceleration can take place. In this subsection, 
we present the results of a numerical evaluation of the achievable 
maximum energies of protons in both stages within the conceivable 
parameter region. 
In Fig.\ref{fig:fig1}, the energies of accelerated protons are plotted 
against the transverse filament size parametrizing the ratio of 
the mean/fluctuate magnetic energy density, $b$. 
The transverse filament sizes, $\lambda$, covers the range from 
$10^{-4}$ pc to $0.067$ pc. The minimum and the maximum sizes 
are compared to the correlation length of Alfvenic fluctuation 
and the jet diameter, respectively. As shown, 
in the case of preliminary acceleration the achievable maximum 
energies of protons are mainly restricted by the escape from 
individual filaments, which is represented by a linearly increasing 
line ($\propto\lambda^{2.65}$).
In the case of $b=10^{-2}$, the proton energies achieved via preliminary 
acceleration, $E_{\rm p}^{*}$, are determined by $r_{\rm g,p}\leq\lambda$ 
over the whole range of the considered filament sizes. In the case of 
$b=10^{-3}$, the values of $E_{\rm p}^{*}$ are also determined by 
$r_{\rm g,p}\leq\lambda$ for $\lambda< 1.7\times 10^{-2}$ pc, while the 
$E_{\rm p}^{*}$-values are more severely restricted by the propagation 
time of the shock for $\lambda\geq 1.7\times 10^{-2}$ pc, and hence the 
achievable energy via the preliminary acceleration,  
$E_{\rm p}^{*}$, is cut-off at $10^{16}$ eV. If $b\leq 10^{-4}$, the  
$E_{\rm p}^{*}$-values are perfectly overlapped with those of $b=10^{-2}$. 
On the other hand, the maximum energy achieved by the additional 
acceleration $E_{\rm p}^{\rm max}$ is independent of not only the 
filament size, but also the $b$-parameter.  
The energies of accelerated protons can be additionally boosted up to 
the value determined by the spatial limit expressed with a horizontal 
line of $E_{\rm p}^{\rm max}=2.1\times 10^{18}$ eV,  
irrespective of their energies acquired by the preliminary 
acceleration. 

\begin{figure}
  \begin{center}
    \FigureFile(80mm,80mm){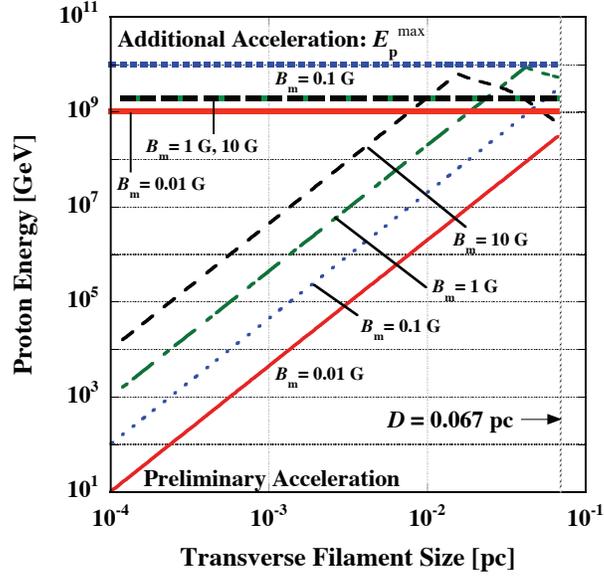}
  \end{center}
  \caption{Maximum proton energies versus the transverse filament 
sizes for various $B_{\rm m}$-values. The value of $b$ is fixed at 
$10^{-3}$. Four monotonically increasing lines 
denote the limitations obtained by preliminary acceleration. The cases of 
$B_{\rm m}=0.01$ G, 0.1 G, 1 G, and 10 G are plotted by solid, dotted, 
dot-dashed, and dashed curves, respectively. Horizontal thick lines 
represent the cutoff energies achieved by additional acceleration. 
The line types are the same as those for preliminary acceleration.}
\label{fig:fig2}
\end{figure}

We also present the $\lambda$-dependence of $E_{\rm p}^{*}$ 
parametrizing the $B_{\rm m}$-value in Fig.\ref{fig:fig2}. The lines 
corresponding to $r_{\rm g,p}=\lambda$ shift upwards with a larger 
$B_{\rm m}$. For the cases of $B_{\rm m}=0.01$ G and 0.1 G, the 
proton energies achieved by the preliminary acceleration, $E_{\rm p}^{*}$,  
are determined by $r_{\rm g,p}\leq\lambda$ over the whole range of 
conceivable filament sizes. In the cases of $B_{\rm m}=1$ G and 10 G, 
the protons dominantly lose their energies via collisions with photons 
for $\lambda\geq 4.2\times 10^{-2}$ pc and for 
$\lambda\geq 1.5\times 10^{-2}$ pc,  respectively. Thus, the resultant 
maximum value of $E_{\rm p}^{*}$ is nearly $10^{19}$ eV. As for the 
additional acceleration, the $E_{\rm p}^{\rm max}$-values are expressed 
by horizontal lines because of their independence of $\lambda$. The 
lowest and the highest $E_{\rm p}^{\rm max}$ correspond to those of 
$B_{\rm m}=10^{-2}$ G and $B_{\rm m}=10^{-1}$ G, respectively, and 
in both cases the maximum energies are restricted by the spatial limit. 
The $E_{\rm p}^{\rm max}$-value for $B_{\rm m}=1$ G is identical to 
that for $B_{\rm m}=10$ G, and in these cases the energy is 
restricted by collisions with ambient photons. 
   
\section{Estimation of the Neutrino Flux from Mrk 501}
In the previous section we calculated the achievable maximum energies 
of protons in a blazar jet. Since the proton energies accelerated by 
multi-stage DSA are sufficiently higher than the neutrino-producing 
threshold via photopionization, we can safely expect significant fluxes 
of energetic neutrinos. 

\subsection{Neutrino Energies Produced by Accelerated Protons}
In the photomeson process of 
$p\gamma\rightarrow\Delta^{++}\rightarrow\pi^{+}N$, the center-of-mass 
energies of accelerated protons should be beyond the $\Delta$-resonance 
threshold, $m_{\Delta}=1230$ MeV,
\begin{equation}
E_{\rm p}^{\rm min}=\Gamma_{\rm J}^{2}\frac{m_{\Delta}^{2}-m_{\rm p}^{2}}
{4E_{\gamma}}, 
\label{eqn:epmin}
\end{equation}
where $\Gamma_{\rm J}$ and $E_{\gamma}$ are the Lorentz factor of a 
blazar jet and the mean energy of ambient photons, respectively. 
Presuming that the generated pion energy is equally divided into four 
leptons through the decay $\pi^{+}\rightarrow\nu_{\mu}\mu^{+}
\rightarrow\nu_{\mu}e^{+}\nu_{e}\bar{\nu_{e}}$, the 
neutrino energy is described as 
\begin{equation}
E_{\nu}=\frac{1}{4}\langle x_{p\rightarrow\pi} \rangle E_{\rm p},
\end{equation}
where $\langle x_{p\rightarrow\pi}\rangle\simeq 0.2$ is the averaged 
fraction of momentum transfer from a proton to a pion.

As for the Mrk 501, the peak energy of synchrotron emssion in the 
X-ray band is shifted from 1 keV to 100 keV during flaring 
(\cite{kataoka01}). Since these energies are achievable for 
synchrotron photons emitted from co-accelerated electrons, we take 
these values as the mean energy of ambient photons. 
For $E_{\gamma}=100$ keV, we obtain the neutrino producing threshold 
using eq.(\ref{eqn:epmin}) as 
\begin{equation}
E_{\rm p}^{\rm min}=1.4\left(\frac{\Gamma_{\rm J}}{30}\right)^{2}
\left(\frac{100~{\rm keV}}{E_{\gamma}}\right)~{\rm PeV},
\end{equation}
which leads to $E_{\nu}^{\rm min}=70$ TeV. Similarly, for $E_{\gamma}
=1$ keV, $E_{\rm p}^{\rm min}=140$ PeV and hence $E_{\nu}^{\rm min}
=7$ PeV. 
Since the maximum energy of accelerated protons is sufficiently higher than 
these threshold values, Mrk 501 can be a feasible source of neutrinos. 
In both cases we determined $E_{\nu}^{\rm max}\simeq 110$ PeV, according 
to the upper limit estimated in \S 2.3. 

\subsection{Energetic Neutrino Flux from Mrk 501}
In order to calculate the neutrino flux produced at an individual blazar, 
we begin with the following useful formula:  
\begin{equation}
\int_{E_{\nu}^{\rm min}}^{E_{\nu}^{\rm max}}E_{\nu}\frac{d\Phi}
{dE_{\nu}}dE_{\nu}=\frac{L_{\nu}}{4\pi d_{L}^{2}},
\end{equation}
where the upper and lower limits of integration are the maximum and 
minimum (threshold) neutrino energies, estimated in \S 3.1, 
respectively. The luminosity distance to the blazar, $d_{L}$, is 
defined as 
\begin{equation}
d_{\rm L}=d_{\rm m}(1+z)=\frac{c(1+z)}{H_{0}}\int_{0}^{z}\sqrt{(1+z)^{2}
(1+\Omega_{\rm m}z)-z(2+z)\Omega_{\Lambda}}dz,
\end{equation}
where $z$ is the redshift (=0.034 for Mrk 501: \cite{quinn96}) for 
Mrk 501 and $d_{\rm m}$ is the proper motion distance, which is 
written by the Hubble constant, $H_{0}=71$ km$^{-1}$Mpc$^{-1}$, the 
matter density, $\Omega_{\rm m}=0.27\pm 0.04$, and dark energy density,  
$\Omega_{\Lambda}=0.73\pm 0.04$, adopted from WMAP results 
(\cite{bennett03}). 

The observed neutrino luminosity, $L_{\nu}$, is defined as 
\begin{equation}
L_{\nu}=\frac{N_{\nu}\langle E_{\nu} \rangle}{\Delta t_{\rm obs}},
\end{equation}
where $N_{\nu}$, $\langle E_{\nu} \rangle$, and $\Delta t_{\rm obs}
\simeq 1~{\rm d}$ are the number of produced neutrinos, the mean 
neutrino energy, and the observed variability time of the emission 
region, respectively. Since $N_{\nu}$ and $\langle E_{\nu} \rangle$ 
are not observables, we introduce the optical depth defined as 
\begin{equation}
\tau=Rn_{\gamma}\sigma_{p\gamma\rightarrow\Delta}=R\frac{L_{\gamma}\Delta
t_{\rm obs}}{V\langle E_{\gamma}\rangle}\sigma_{p\gamma\rightarrow \Delta},
\end{equation} 
where $L_{\gamma}\simeq 10^{45}$ erg/s is the observed photon luminosity, 
$V$ represents the volume of emission region with radius $R$, 
$\langle E_{\gamma}\rangle\simeq 10$ eV denotes the mean energy of photons, 
and $\sigma_{p\gamma\rightarrow\Delta}\simeq 10^{-28}$ cm$^{2}$ is the 
$\Delta$-resonance photopionization cross section. Then, we can express 
$L_{\nu}$ with observable quantities in a compact notation, 
\begin{equation}
L_{\nu}=K\tau L_{\rm p}=K\tau e^{\tau}L_{\rm p,obs},
\end{equation}
where $K\simeq 0.024$ is a constant taking the branching ratios of 
interaction chain and corresponding momentum transfer into account. 
The intrinsic proton luminosity, $L_{\rm p}$, can be replaced with the 
observed $L_{\rm p,obs}$ taking the proton interactions with ambient 
photons into consideration. However, $L_{\rm p,obs}$ is still unknown; 
we assume it to be as 10 $\%$ of the photon total luminosity 
(\cite{halzen02}). We also replace $\tau$ with $(1-e^{-\tau})$, with 
considering the possible p and $\pi^{+}$ absorption in the vicinity of 
the source. 

Summarizing the above discussion, we adopt the following formula of 
differential flux from an individual blazar (\cite{bazo05}):  
\begin{equation}
\frac{d\Phi_{\nu_{\mu}+{\overline\nu_{\mu}}}}{dE_{\nu}}=
\frac{2\frac{0.1KL_{\gamma}(1-e^{-\tau})e^{(1-e^{-\tau})}}
{4\pi d_{L}^{2}}}
{\int_{E_{\nu}^{\rm min}}^{E_{\nu}^{\rm max}}E_{\nu}E_{\nu}^{-p}
\exp\left(-\frac{E_{\nu}}{E_{\rm cut}^{\rm AGN}}dE_{\nu}\right)}
E_{\nu}^{-p}
\exp\left(-\frac{E_{\nu}}{E_{\rm cut}^{\rm AGN}}\right),
\label{eqn:dflux}
\end{equation}
where the factor 2 reflects the contribution of neutrinos and 
anti-neutrinos.
Since we are concerned with the point-source neutrino flux, we 
consider the $\nu_{\mu}$ contribution on account of their highest 
angular resolution. This is attributed to the large muon track, 
having advantege of reconstructing paths.  

\begin{figure}
  \begin{center}
    \FigureFile(80mm,80mm){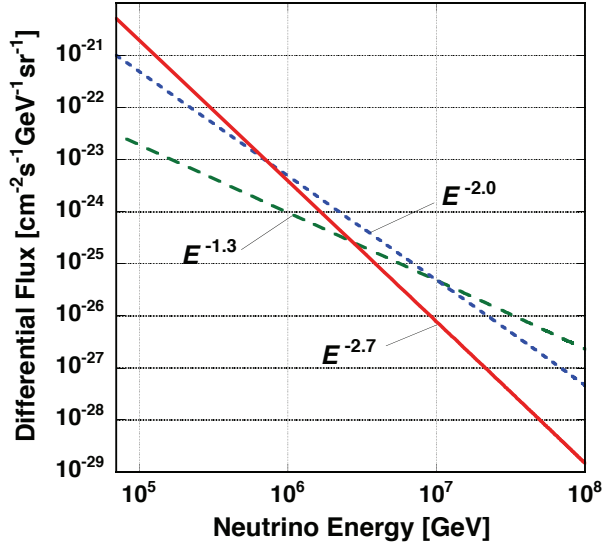}
  \end{center}
  \caption{Differential neutrino flux from a blazar jet of Mrk 501 
versus neutrino energy. The power-law energy distribution of neutrinos,  
$E_{\nu}^{-p}$, with an exponential cutoff $E_{\rm cut}^{\rm AGN}=2.1
\times 10^{18}$ eV is assumed. The power-law indices of $p=1.3$, 2.0, 2.7 
are denoted by the dashed, dotted, and solid curves, respectively.}
\label{fig:fig3}
\end{figure}

In Fig.\ref{fig:fig3} we plot the differential flux of neutrinos 
from Mrk 501 against their energies. The energy of neutrinos is 
limited to the range between the maximum and the minimum predicted 
by this model. Here, we adopt the typical 
power-law energy distribution of neutrinos, $E_{\nu}^{-p}$, with 
the exponential cutoff, $E_{\rm cut}^{\rm AGN}$. We parametrize 
the power-law index $p=1.3$, 2.0, 2.7, where $p=1.3$ and $p=2.7$ 
are taken from the upper and lower limits of the exponents, 
respectively, and $p=2.0$ is a resemblance of the parent proton spectrum 
supposed to be accelerated by Fermi's mechanism. The cutoff energy,  
$E_{\rm cut}^{\rm AGN}=2.1\times 10^{9}$ GeV, is the achievable 
maximum energy of AGN protons estimated in $\S 2.3$.
One can see that for a steeper energy spectrum the neutrino flux is 
more significantly reduced in the energy range of $E_{\nu}\gtrsim 10^{8}$ 
GeV. On the contrary, in the lower energy region of 
$E_{\nu}\lesssim 10^{6}$ GeV the flux is higher in the case of a 
steeper spectrum because of the contribution from the denominator. 
Anyhow, the neutrino flux originating from a single blazar is trivial 
compared to the observable level. 

\subsection{Calculation of the $\nu_{\mu}$-Induced Event Number}
It is more difficult to detect higher energy neutrinos because of the 
steeply falling spectrum. In order to detect the interaction of a TeV 
neutrino with a Cherenkov telescope, the effective volume of the 
detector is required for kilometer scale to cover the typical range 
of TeV muons. In the effective region of the telescope, the 
probability to detect $\nu_{\mu}$ in the TeV-PeV range is 
approximately given by 
\begin{equation}
P_{\nu_{\mu\rightarrow}\mu}\simeq \frac{R_{\mu}}{\lambda_{\rm int}}
\simeq 1.3\times 10^{-6}\left(\frac{E_{\nu}}
{1~{\rm TeV}}\right)^{0.8},
\label{eqn:prob}
\end{equation}
where $R_{\mu}$ and $\lambda_{\rm int}$ are the muon range and the 
neutrino interaction length, respectively (\cite{gaisser95}).   

We can now compute the diffuse neutrino event rate by integrating 
the total differential flux multiplied by the detection probability 
of eq.(\ref{eqn:prob})   
\begin{equation}
N_{\nu_{\mu}}=\int_{E_{\rm \nu}^{\rm min}}^{E_{\rm \nu}^{\rm max}}
dE_{\nu_{\mu}}\left<\frac{d\Phi_{\nu_{\mu}}}{dE_{\nu_{\mu}}}
(E_{\nu_{\mu}})\right>_{\rm tot}P_{\nu_{\mu}\rightarrow\mu}, 
\end{equation} 
where the upper and lower limits of the integral are 
$E_{\rm \nu}^{\rm max}\simeq 1.1\times 10^{17}$ eV and $E_{\rm \nu}
^{\rm min}\simeq 7\times 10^{13}$ eV calculated in $\S$ 3.1, 
respectively. In order to compute the diffuse neutrino flux from 
the observed blazar distribution, we should integrate the flux from 
all blazars while taking the Doppler factor distribution into 
consideration. For simplicity, here we adopt the effective number of 
blazars (\cite{halzen97}), using the $\gamma$-ray flux ratio of 
diffusive to the single blazar, which is derived from the luminosity 
function of 20 brightest blazars obtained by the EGRET 
collaboration (\cite{chiang95}). Thus, instead of summing up the total 
isotropic differential flux of neutrinos 
$\left<\frac{d\Phi_{\nu_{\mu}}}{dE_{\nu_{\mu}}}\right>_{\rm tot}$,  
we simply estimate the diffuse flux by multiplying the 
flux of Mrk 501 by the resultant number 130 sr$^{-1}$. A correction 
for the difference in the spectral indices of gamma-ray and neutrino fluxes 
enhances the neutrino flux by a factor of three. 
We present the estimated neutrino event rates in Table 1 for conceivable 
power-law exponents. Considering the 4$\pi$ coverage and typical exposure 
time ($\sim 10$ yr), the value $N_{\nu_{\mu}}\simeq 2$ km$^{-2}$ yr$^{-1}$ 
sufficiently reaches to the observable level.  

\begin{table}[h]
\begin{center}
\caption{Neutrino event rates for various power-law exponents.\label{tbl-1}}
\begin{tabular}{|c|c|}\hline
Power-law exponent & Event rate \\ \hline
$p=1.3$ & $0.78$ \\ \hline
$p=2.0$ & $1.3$ \\ \hline
$p=2.7$ & $1.9$ \\ \hline
\end{tabular}
\end{center}
\end{table}
   
\section{Conclusions}
In this paper we applied our model of multi-stage diffusive shock 
acceleration to the blazar Mrk 501 and evaluated the maximum energies 
of protons. We also calculated the neutrino flux from this single 
source generated by these accelerated protons, and discuss the 
detectability with the updated neutrino telescope, IceCube. We 
obtained the EeV protons and differential flux during the operating time 
of the detector by not a simple, optimistic order estimation, but by 
considering the systematical energy restriction at the source. The 
results are rather sensitive to the values of the magnetic field 
strength and the size of the accelerator. 
The $B$-field value is related to the system size via the radiation 
intensity of the emission region. In the conceivable range of the field 
strength, EeV protons are obtained by additional acceleration. 
 
\bigskip
Y. S. H. thanks K. Murase for useful comments.

\end{document}